\documentclass[aps,twocolumn,prc,superscriptaddress,showpacs,nofootinbib,floatfix,amssymb,amsfonts,amsmath]{revtex4-1}

\usepackage{graphicx}
\usepackage{epsf}
\usepackage{amsmath,amssymb}
\usepackage{bm}
\usepackage{color}
\usepackage{ulem}

\begin{document}

\title{
The possibility of $^{14}$C cluster as a building block of medium mass nuclei
}%

\author{N. Itagaki}
\affiliation{
Yukawa Institute for Theoretical Physics, Kyoto University,
Kitashirakawa Oiwake-Cho, Kyoto 606-8502, Japan
}

\author{A. V. Afanasjev}
\affiliation{
Yukawa Institute for Theoretical Physics, Kyoto University,
Kitashirakawa Oiwake-Cho, Kyoto 606-8502, Japan
}
\affiliation{
Department of Physics and Astronomy, Mississippi State University, MS 39762, USA
}

\author{D. Ray}
\affiliation{
Institute for Systems Engineering Research (ISER), Mississippi State University, MS 39762, USA
}

\date{\today}

\begin{abstract}
The possibility of the $^{14}$C cluster being a basic building block 
of medium mass nuclei is discussed.
 Although $\alpha$ cluster structures have been widely discussed in 
the light $N\approx Z$ mass region,  
the neutron to proton ratio deviates from unity in 
the nuclei near $\beta$-stability line and in neutron-rich nuclei. Thus, more neutron-rich objects with
$N>Z$ could become the building blocks of cluster structures in such nuclei.
The $^{14}$C nucleus is 
strongly bound and can be regarded as such a candidate. In addition, the path 
to the lowest shell-model configuration at short relative distances is closed for the $^{14}$C+$^{14}$C 
structure contrary to the case of the $^{12}$C+$^{12}$C structure; this allows to keep appreciable 
separation distance  between the $^{14}$C clusters.
The recent development of antisymmetrized quasi-cluster model (AQCM) allows us to utilize $jj$-coupling 
shell model wave function for each cluster in a simplified way. The AQCM results for the $^{14}$C+$^{14}$C
structure in $^{28}$Mg are  compared with the ones of cranked relativistic mean field (CRMF) calculations.
Although  theoretical frameworks of these two models are 
quite different, they give similar results for the nucleonic densities and rotational properties of the structure under 
investigation.
The existence of linear chain three $^{14}$C cluster structure in $^{42}$Ar 
has  also been predicted in AQCM.
These results confirm the role of the $^{14}$C cluster as a possible building block of cluster structures in 
medium mass nuclei. 
\end{abstract}

\maketitle


The $\alpha$ cluster structures have
been extensively studied over the years~\cite{Brink,RevModPhys.90.035004}. 
Since the binding energy per nucleon  
is extremely large in $^4$He, it can be a building block
of the nuclear systems
called $\alpha$ cluster. Also, the relative interaction between the $\alpha$ 
clusters is weak, which is another condition for the appearance of the cluster 
structures. For example,
$^8$Be is not bound but its 
ground state has a developed $\alpha$+$\alpha$ cluster structure.
The candidates for  $\alpha$ cluster states have been widely discussed 
in other light $4N$ ($N$ is integer here) nuclei~\cite{PTPS.68.29}.
This is illustrated by a few examples below.
The second $0^+$ state of $^{12}$C
at $E_x = 7.65$~MeV, located
just above the threshold energy to the decay into three $\alpha$'s,
has a developed three-$\alpha$ cluster structure 
and plays a crucial role in
the formation of  $^{12}$C in stars~\cite{Hoyle}.
In $^{16}$O, the first excited state at $E_x = 6.05$~MeV, located very close to the
threshold of the decay into $^{12}$C and $^4$He,
can be interpreted as $^{12}$C+$^4$He cluster state.
It has been known as the mysterious $0^+$ state;
the reproduction of this state based on the standard picture (shell-model
approaches) is still a big challenge. Various cluster structures have been proposed 
also in $^{20}$Ne, $^{24}$Mg, and $^{44}$Ti $etc.$~\cite{,Brink,RevModPhys.90.035004,PTPS.68.29}.

The description of such cluster structures has been attempted
in simple cluster models. However, 
it is well known that non-central nuclear  interactions are very important
in nuclear systems  and their effects cannot be taken into account in simple $\alpha$ cluster models.
With increasing mass number, the symmetry of the $jj$-coupling shell model
dominates the nuclear structure,
and subclosure configurations of
$j$-shells,  
$f_{7/2}$, $g_{9/2}$, and $h_{11/2}$ become important, 
corresponding to the magic numbers of 28, 50, and 126~\cite{Mayer}. 
Indeed the observation of these magic numbers is the evidence that the spin-orbit interaction
strongly contributes in the medium and heavy mass regions,
and this  interaction is known to play a substantial role in breaking the
$\alpha$ clusters~\cite{PhysRevC.70.054307}.

Therefore, it is natural to think about a different object as a cluster
in the study of medium mass nuclei. 
Here one should also consider that neutron to proton ratio
of stable nuclei deviates from unity with increasing mass number.
Thus, there is the possibility that
more neutron-rich object could be a building block of cluster structures.
In this study, we discuss the possibility that
the $^{14}$C nucleus could be a cluster. This can be justified 
by the following arguments.

First of all, $^{14}$C is strongly bound. 
This is because the proton number 6 corresponds to the subclosure of 
the $p_{3/2}$ subshell of the $jj$-coupling shell model, and the neutron number 8 
is the magic number corresponding to the closure of the $p$ shell.
Although $^{14}$C has two valence neutrons, the lowest threshold to emit 
particle is the neutron threshold at $E_x = 8.18$~MeV, which is high 
enough value. In addition, because of strong shell effects, 
there is no excited state below $E_x = 6$~MeV just like in $^{16}$O.
The $\beta$-decay of free $^{14}$C is very slow reflecting the stability of this nucleus.
Thus, $^{14}$C is a famous nucleus 
used for the age determination~\cite{age-determination}.
The second argument is the following:  although the single $^{14}$C nucleus
is $\beta$-unstable, the line connecting the origin of the nuclear landscape
and the point of 6 protons and 8 neutrons (corresponding to $^{14}$C) on the nuclear chart,
which has $N/Z \sim 1.3$, extrapolates into $\beta$-stability line
above $Z\approx 40$.  
 The third argument is that it is well known that $^{14}$C is emitted from 
some of heavy nuclei [such as $^{221-224,226}$Ra, $^{223,225}$Ac and $^{221}$Fr
(see Ref.\ \cite{Guglielmetti_2008})]  in the process which is called cluster decay
~\cite{cluster-decay}. In reality, there is much more experimental data for 
the $^{14}$C emission as compared with the emission of $^{12}$C in medium and heavy mass
nuclei,  and neutron richness 
of $^{14}$C can be important for that. These experimental data also strongly suggest
that  $^{14}$C can be a building block of cluster structures in heavy nuclei.
Finally, contrary to the case of the $^{12}$C+$^{12}$C cluster structure, the 
 path to the lowest shell-model configuration is closed at short relative distances
 in the $^{14}$C+$^{14}$C cluster structure. This factor leads to appreciable
 distance between the $^{14}$C clusters and it is a subject of the present study.

In this Letter, the appearance of $^{14}$C cluster structures is investigated and 
the predictions obtained in the cluster and mean field approaches
are compared.
The protons of $^{14}$C correspond to the subclosure of $p_{3/2}$ in
the $jj$-coupling shell model.   The $jj$-coupling shell model wave functions can be easily prepared 
starting from the 
cluster model; in Refs. ~\cite{PhysRevC.94.064324,PhysRevC.73.034310,PhysRevC.75.054309,PhysRevC.79.034308,PhysRevC.83.014302,PhysRevC.87.054334,ptep093D01,ptep063D01,ptepptx161,PhysRevC.97.014307,PhysRevC.98.044306}, 
the antisymmetrized quasi-cluster model (AQCM), which allows to smoothly transform $\alpha$ cluster model 
wave functions to $jj$-coupling shell model ones and incorporate the effects of the spin-orbit interaction, has been 
proposed.  
A reliable nucleon-nucleon interaction, which includes both the cluster and shell features in the light 
and medium mass nuclei, is inevitably needed. 
The Tohsaki interaction, which has finite range three-body 
terms~\cite{PhysRevC.49.1814,PhysRevC.94.064324,PTP.94.1019,PhysRevC.98.044306},
is employed here.
Although this is a phenomenological interaction, it provides a reasonable size and 
binding energy for the $\alpha$ cluster and reproduces $\alpha$+$\alpha$ scattering 
phase shifts. In addition, it describes the saturation properties of nuclear matter rather
well.


In AQCM, each single-particle wave function is described by a Gaussian,
\begin{equation}	
	\phi = \left(  \frac{2\nu}{ \pi } \right)^{\frac{3}{4}} 
		\exp \left[-  \nu \left(\bm{r} - \bm{\zeta} \right)^{2} \right] \chi, 
\label{spwf} 
\end{equation}
where the Gaussian center parameter $\bm{\zeta}$ shows the expectation 
value of the position of the particle,
and $\chi$ is the spin-isospin wave function.
The size parameter $\nu$ is set to 0.17~fm$^{-2}$ for $^{28}$Mg (two $^{14}$C) and $^{42}$Ar (three $^{14}$C).
The Slater determinant in the conventional Brink model~\cite{Brink} is constructed from 
these single particle wave functions by antisymmetrizing them. 
Here, four single particle 
wave functions with different spin and isospin
sharing a common Gaussian center parameter
$\bm{\zeta}$ correspond to an $\alpha$ cluster.
In the conventional cluster models,
there is no spin-orbit effect for the $\alpha$ clusters.
Thus, they are changed into quasi-clusters based on 
AQCM~\cite{PhysRevC.94.064324,PhysRevC.73.034310,PhysRevC.75.054309,
PhysRevC.79.034308,PhysRevC.83.014302,PhysRevC.87.054334,
ptep093D01,ptep063D01,ptepptx161,PhysRevC.97.014307}.
According to AQCM,
when the original position of one of the particles
(the value of Gaussian center parameter) is $\bm{R}$,
the Gaussian center parameter of this nucleon is transformed 
by adding the imaginary part as
\begin{equation}
\bm{\zeta} = \bm{R} + i \Lambda \bm{e}^{\text{spin}} \times \bm{R}, 
\label{AQCM}
\end{equation}
where $\bm{e}^{\text{spin}}$ is a unit vector for the intrinsic-spin orientation of this
nucleon. It has been previously shown that
the lowest configurations of the $jj$-coupling shell model
can be achieved by $\Lambda = 1$ and $\bm{R} \to 0$ for all the nucleons~\cite{ptep093D01}.

For the description of $^{14}$C,  at first, di-nucleon clusters are prepared;
in each di-nucleon cluster, two nucleons with opposite spin and same isospin are
sharing common Gaussian center parameters.
Four di-nucleon clusters with a tetrahedron configuration
(the distance between two di-neutron clusters is parameterized as $R$)
and small relative distances ($R \to 0$) corresponds to the closure of the $p$-shell,
which is introduced for the neutron part. 
In the calculations, $R$ is set to 0.1~fm.
For the proton part,
three di-proton clusters with equilateral triangular configuration and small
distance between them are introduced, and 
the imaginary parts of the Gaussian center parameters are given
as $\Lambda = 1$ in Eq.~(\ref{AQCM}),
which correspond to the subclosure of the $p_{3/2}$ shell~\cite{PhysRevC.94.064324,PhysRevC.87.054334}.

For the analysis based on AQCM, the Hamiltonian consists of kinetic energy and 
potential energy terms, and the potential energy has central, spin-orbit, and 
Coulomb parts. For the central part, the Tohsaki interaction~\cite{PhysRevC.49.1814}  
is adopted, which has finite range three-body nucleon-nucleon interaction terms in 
addition to two-body terms. This interaction is designed to reproduce both saturation 
properties and scattering phase shifts of two $\alpha$ clusters. For the spin-orbit part, 
the spin-orbit term of  the G3RS interaction~\cite{PTP.39.91}, which is a realistic 
interaction originally  developed to reproduce the nucleon-nucleon scattering phase 
shifts, is adopted. The combination of these two has been well 
investigated~\cite{PhysRevC.97.014307,PhysRevC.98.044306}.

\begin{figure}[htbp]
	\centering
	\includegraphics[width=6.5cm]{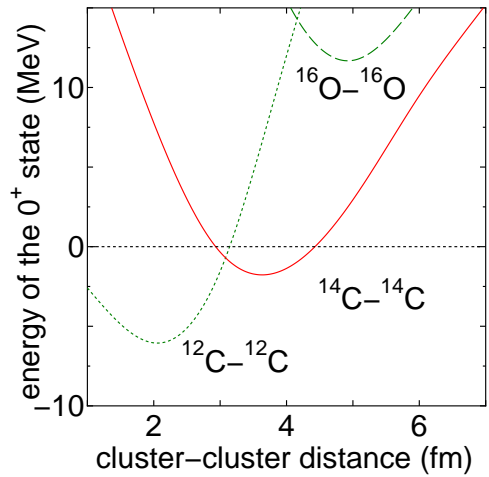} 
	\caption{
The $0^+$ energy curves of 
$^{12}$C+$^{12}$C ($^{24}$Mg, dotted line)
and $^{14}$C+$^{14}$C ($^{28}$Mg, solid line) measured from the
threshold energies.
The dashed curve corresponds to the $0^+$ energy curve of $^{16}$O+$^{16}$O ($^{32}$S).
 The threshold energies are estimated by adopting the optimal $R$ and $\Lambda$ values
 for the free $^{12}$C, $^{14}$C, and $^{16}$O nuclei with the size parameter of $\nu = 0.2$~fm$^{-2}$.
}
\label{c-c-energy}
\end{figure}

We start the discussion with the comparison of the 
$^{12}$C+$^{12}$C and $^{14}$C+$^{14}$C 
systems. 
The energy curves of the $0^+$ states of these two systems measured from the threshold 
energies are shown in Fig.~\ref{c-c-energy}. 
It is confirmed
that the minimum energy is much lower 
than the threshold in the case of the
$^{12}$C+$^{12}$C system (dotted line). 
The optimal distance is rather small, around 2~fm.
According to the $jj$-coupling shell model, the 
small distance limit of two $^{12}$C corresponds to the lowest configuration of $^{24}$Mg.
This means that the path going to the ground state after the fusion is opened for 
the $^{12}$C+$^{12}$C system.
This situation is different in the case of
$^{14}$C+$^{14}$C system (solid line).
Small distance limit of two $^{14}$C does not correspond to the lowest 
$jj$-coupling shell model wave function of $^{28}$Mg. As a result, the minimum energy
appears somewhat below the threshold.
This result suggests the possibility that $^{14}$C clusters keep the relative distance 
($\sim$4~fm), which is much larger than in the case of the $^{12}$C+$^{12}$C system,
and form a developed cluster structure.
The dashed line is for the energy of the
$^{16}$O+$^{16}$O system measured from the threshold.
In this case, again the path going to the ground state is closed,
which works for the appearance of a well-developed cluster structure.
However, the energy is much higher than the threshold 
because of large Coulomb repulsion.
The  $^{16}$O+$^{16}$O structure of $^{32}$S has been discussed for years~\cite{PhysRevC.66.021301},
but it has been known that the $^{16}$O+$^{16}$O cluster component corresponds to 
highly excited states~\cite{PhysRevC.69.051304}. 
The possibility of the existence of highly excited
superdeformed bands in $^{32}$S based on the $^{16}$O+$^{16}$O structure  has
also been discussed in density functional theories \cite{RER.00,MKKRHT.06,RA.16}.
In the case of $^{14}$C+$^{14}$C system, the state appears around the threshold because of 
smaller Coulomb repulsion.

\begin{figure}[ht]
\centering
\includegraphics[width=8.5cm]{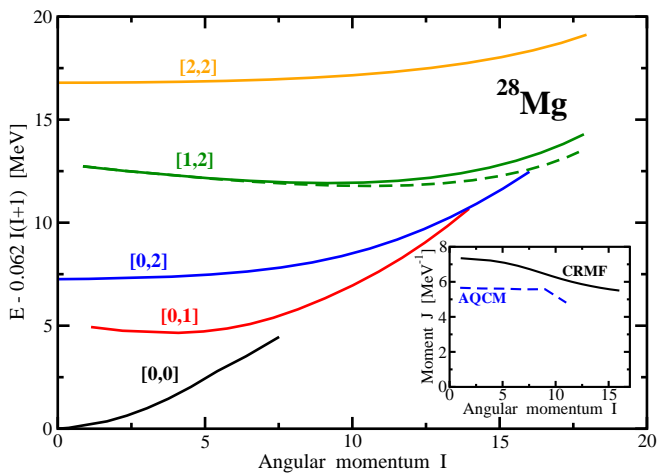}
\caption{Excitation energies of calculated CRMF configurations 
in $^{28}$Mg relative to a rotating liquid drop reference $A I(I + 1)$, 
with the inertia parameter $A = 0.062$.
The insert shows  the moments of inertia of the $^{14}$C+$^{14}$C
cluster structure as a function of the spin.}
\label{E_RLD}
\end{figure}

 An alternative way to look on clustering in nuclei is through the prism of
 density functional theories (DFT). Both relativistic and non-relativistic DFTs 
 have been applied to the investigation of this phenomenon in nuclei (see 
 Refs.~\cite{EKNV.12,RA.16} and references quoted therein).  The advantage of the 
 DFT framework is the fact that it does not assume the existence of cluster structures;
 the formation of cluster structures proceeds from microscopic  single-nucleon 
 degrees of freedom via many-body  correlations~\cite{MARUHN20101,EKNV.12,EKNV.14,AA.18}. 
 As a result, the DFT framework allows the simultaneous treatment of cluster and 
 mean-field-type states. 
 
This is a reason why cranked relativistic mean field (CRMF) approach (see
Ref.~\cite{VALR.05}) is also used in the present manuscript for the study of 
clusterization in $^{28}$Mg. This 
 approach has been successfully employed for the study of both the clusterization 
 in light nuclei ranging from $^{12}$C up to $^{50}$Ti \cite{ZIM.15,RA.16,AA.18,PhysRevLett.107.112501,PhysRevC.90.054307,PhysRevC.92.011303}  
 and of the rotation of light, medium and heavy mass nuclei (see Ref.\ \cite{VALR.05} 
 and references quoted therein). In the CRMF approach the nucleus is described as
 a system of pointlike nucleons, Dirac spinors, coupled to mesons and to the photons
(see Ref.\ \cite{VALR.05}). The nucleons interact by the exchange of several mesons,
namely, a scalar meson $\sigma$ and three vector particles, $\omega$, $\rho$ and
the photon. The CRMF calculations are performed with the NL3* functional~\cite{NL3*}. 
     
\begin{figure}[ht]
\centering
\includegraphics[width=4.2cm]{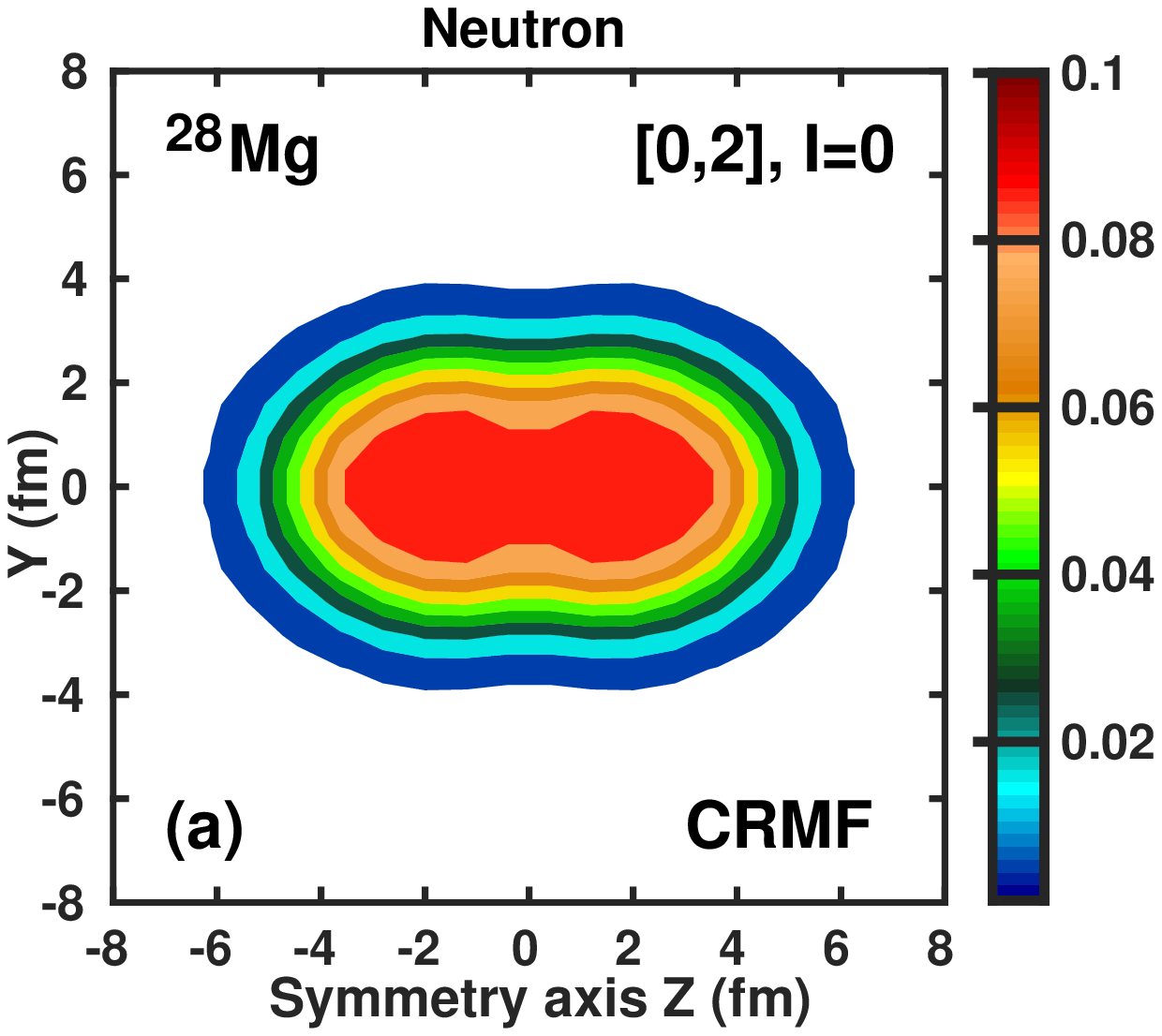}
\includegraphics[width=4.2cm]{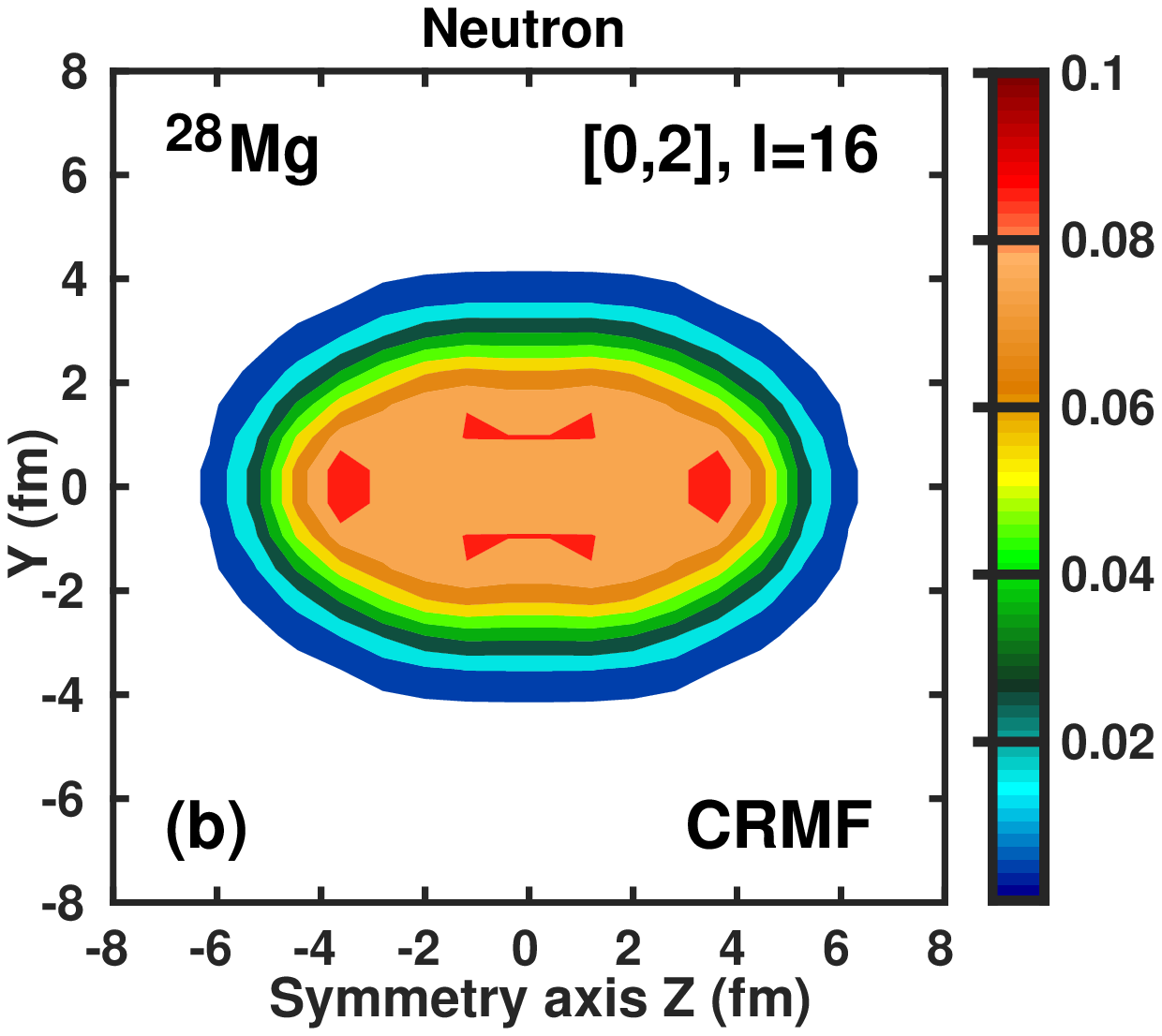}
\includegraphics[width=4.2cm]{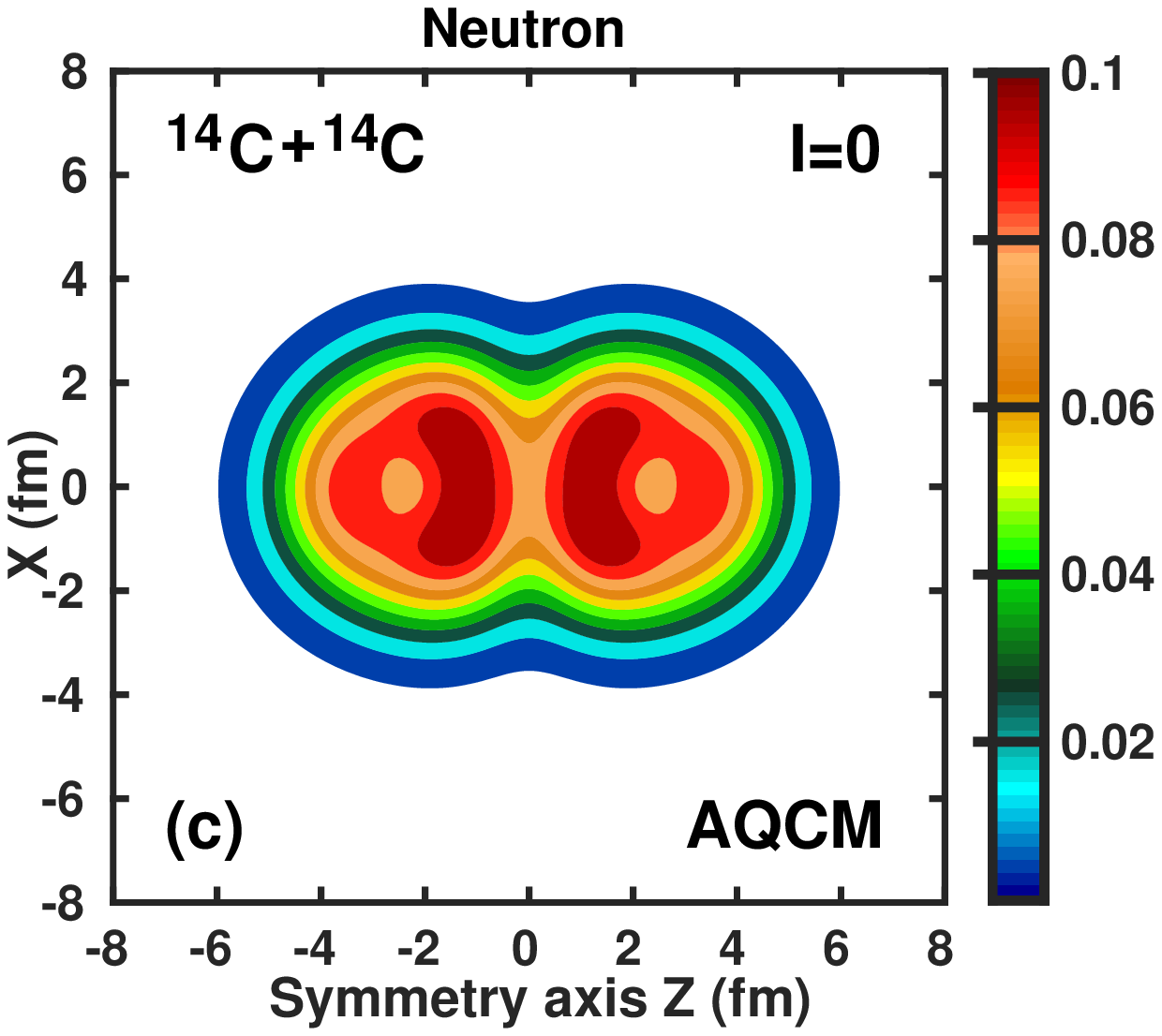}
\includegraphics[width=4.2cm]{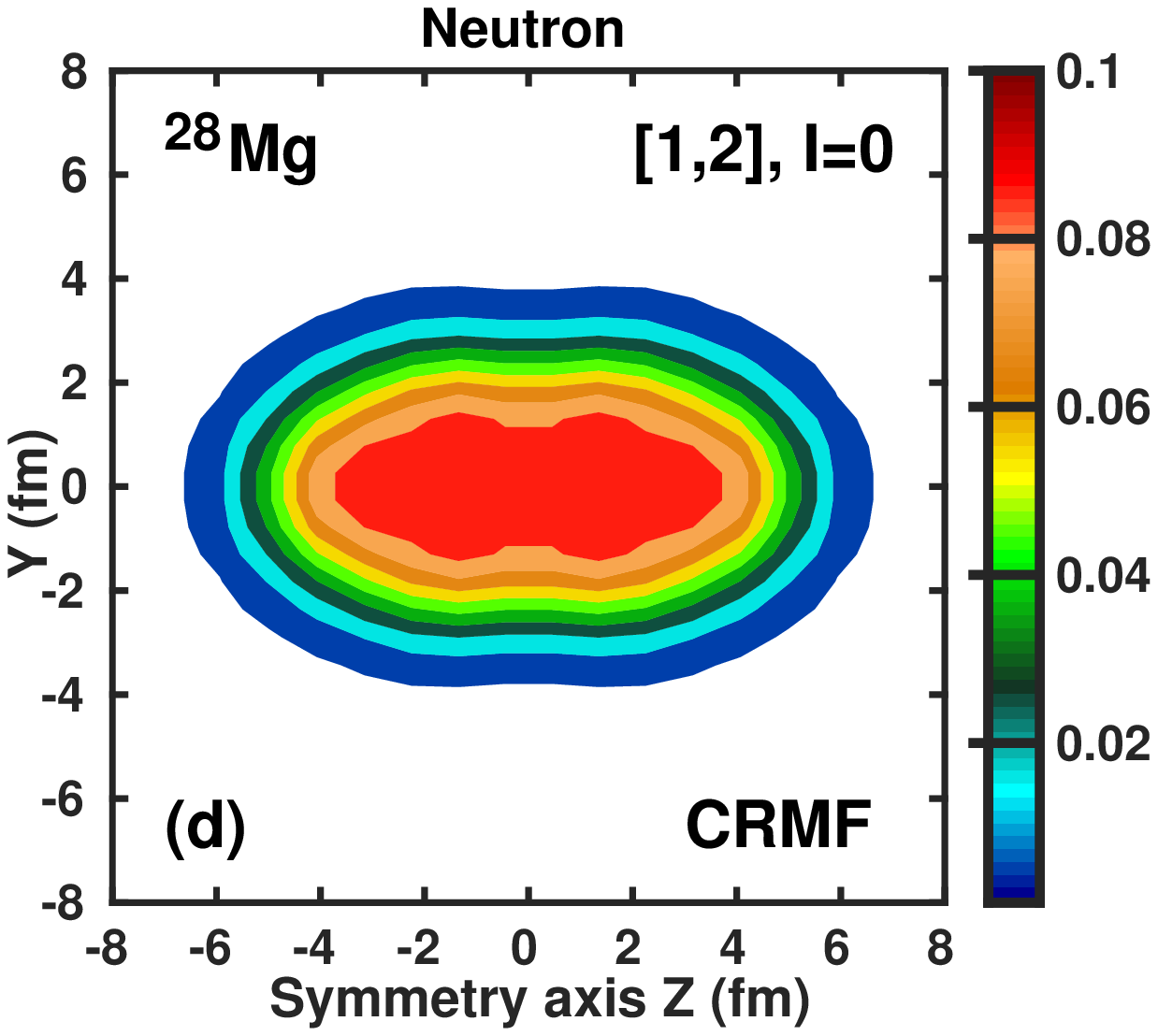}
\caption{Neutron density distributions of the configurations of interest.
Panels (a), (b) and (d) show the densities of the [0,2] and [1,2] configurations 
obtained in the CRMF calculations, while panel (c) the densities of the  
$^{14}$C+$^{14}$C  structure obtained in the AQCM calculations. The density 
colormap starts at $\rho =0.005$ fm${^{-3}}$ and shows the densities in 
fm$^{{-3}}$. }
\label{density}
\end{figure}

 Figure~\ref{E_RLD} shows calculated  energies of the lowest configurations in 
 $^{28}$Mg. The configurations are labeled by shorthand $[p,n]$ labels where 
 $p$ ($n$) is the number of occupied $N=3$ proton (neutron) intruder orbitals
(here $N$ is principal quantum number).
 The ground state band [0,0] has no such orbitals occupied and it has 
 quadrupole deformation of $\beta_2 = 0.34$ at spin $I=0$.  As discussed in Ref.\ 
 \cite{RA.16}, such normal deformed band has limited angular momentum content 
 and it terminates in purely single-particle state at $I=8$. Subsequent particle-hole 
 excitations lead to 
 rotational bands with structure [0,1], [0,2], 
 [1,2], [2,2], ...,  which have larger angular momentum content (see Fig.\ \ref{E_RLD})
 and larger deformation.
 Neutron densities of selected states of these rotational bands are shown in Fig.~\ref{density}.  
The density distribution of the [0,2] configuration at spin $I = 0$ 
(Fig.~\ref{density} (a))
corresponds to the $^{14}$C+$^{14}$C cluster structure and it is similar (especially in 
high-density region) to the one obtained in the  AQCM (Fig.~\ref{density}  (c)). 
In both calculations, the $^{14}$C+$^{14}$C distance is approximately equal to  4~fm.

However, this cluster  structure behaves differently as a function of 
the spin in the CRMF and AQCM calculations. This is illustrated in the insert to 
Fig.~\ref{E_RLD} which shows  the moments of inertia of the $^{14}$C+$^{14}$C
cluster structure as a function of the spin. In  the CRMF calculations, the moment of inertia 
is gradually decreasing with spin. This is due to two factors: the decrease of quadrupole 
deformation and washing out of clusterization with increasing spin which is generally 
observed in DFTs calculations (see Refs.~\cite{RA.16,AA.18}). The comparison of
the densities at spins $I=0$ and 16 illustrates the latter feature (see Figs.~\ref{density}~(a) and (b)).
On the contrary, the moments of inertia are somewhat smaller in the AQCM 
calculations and they stay almost constant
up to $I = 8$.
Note that in AQCM it is assumed that cluster structure persists even at highest
calculated spins.
 In the light of distinct predictions of these two models, the 
experimental observation of the superdeformed band built on the  $^{14}$C+$^{14}$C 
structure would be extremely useful for clarification of existing differences in  the 
description of clusterization in the DFT and cluster-originating models.
 
Note that in the CRMF calculations the [0,2] configuration, representing 
the  $^{14}$C+$^{14}$C cluster structure, plays a role of a basic building 
block of more elongated cluster structures with the [1,2] configurations, 
which  are created by means of particle-hole excitations.  At spin zero, the total (proton 
+ neutron) quadrupole deformations of the [0,2] and [1,2] configurations 
are 0.81 and 0.89, respectively. The density of the [1,2] configuration is 
shown in Fig.~\ref{density}~(d). The differences in the density distributions of the 
[0,2] and [1,2] configurations are due to proton particle-hole excitation from the
3/2[211] orbital into 1/2[330] one. These orbitals have different spatial distributions
of the single-particle density (see Ref.~\cite{AA.18,A-epj-wc.18} for details) and thus such a 
particle-hole excitation moves density from the middle part of the nucleus to the polar 
region leading to a more elongated shape (compare Fig.~\ref{density}~(d)
with Fig.~\ref{density}~(a)). Note that there are two [1,2] configurations
which are signature degenerated up to $I\sim 10$ because of the degeneracies
of the $3/2[211](r=\pm i)$ orbitals. Subsequent particle-hole excitations, 
leading to the [2,2] configuration with quadrupole deformations of 1.13 at spin 
zero, partially suppress $^{14}$C+$^{14}$C cluster structure and form 
ellipsoidal-like  density distribution.



\begin{figure}[htbp]
	\centering
	\includegraphics[width=6.5cm]{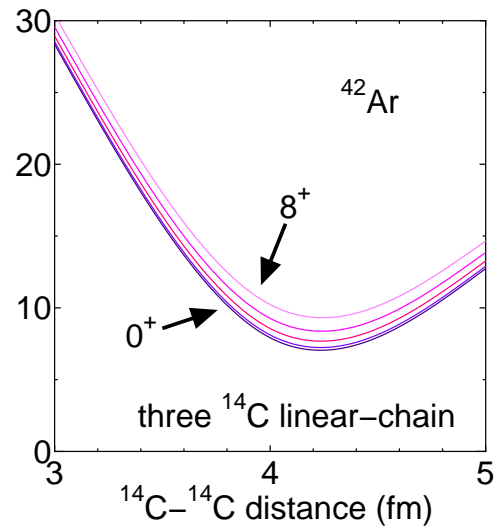} 
	\caption{
The energy curves for  the linear chain of three $^{14}$C clusters ($^{42}$Ar)
as a function of the $^{14}$C-$^{14}$C distance.
The angular momentum is changed from 0 to 8.
 The threshold energies are estimated by adopting the optimal $R$ and $\Lambda$ values
 for free $^{14}$C with the size parameter of $\nu = 0.2$~fm$^{-2}$.
}
\label{three-c-lin}
\end{figure}

Next, we consider the case of linear chain structure of three $^{14}$C clusters 
($^{14}$C+$^{14}$C+$^{14}$C) in $^{42}$Ar. The energy curves for such a
structure are shown in Fig.~\ref{three-c-lin} as a function of the $^{14}$C+$^{14}$C
distance. The AQCM solutions for the states with angular momentum ranging
from $I=0$ up to $I=8$ are shown in this figure.  
Similar to the case of cluster $^{14}$C+$^{14}$C structure 
in $^{28}$Mg, the minima of the energy curves are obtained around the threshold 
energy with the relative distance of the $^{14}$C clusters being approximately 
equal to  $\sim$4~fm. The moment of inertia of
the calculated rotational band is 15.59~MeV$^{-1}$. 
In the future, one should investigate
the stability of the linear chain state against the bending motion.
Even if it is not stable against the bending motion at spin zero, 
there is a possibility that it becomes stabilized when 
large angular momentum is given to the system. This is because
it is well known that rotation can stabilize elongated shapes
owing to the  centrifugal force~\cite{PhysRevLett.107.112501,PhysRevC.90.054307,PhysRevC.92.011303}.

In conclusion, the possibility that $^{14}$C can be a building block of  cluster
structures  in medium mass nuclei has been investigated for the first time.  On
going from light to heavier nuclei, the beta-stability line gradually evolves from
$N\sim Z$ to $N\sim 1.3 Z$.  This suggests that the nuclei with similar neutron to proton
ratio can be the building blocks of cluster structures in medium mass nuclei.  The 
$^{14}$C nucleus is such a candidate since it has $N/Z=1.33$. It is strongly bound and 
reveals itself  as a cluster in the $^{14}$C emission from actinides.
Moreover, the path to the lowest shell-model configuration at short relative distances
is closed in the $^{14}$C+$^{14}$C structure, which allows to keep
the distance between the clusters approximately equal to 4 fm. Despite underlying
differences in theoretical assumptions, the AQCM and CRMF calculations predict the 
existence of such a cluster structure in $^{28}$Mg. In addition, the possible
existence of linear chain $^{14}$C+$^{14}$C+$^{14}$C cluster structure in $^{42}$Ar
has been investigated within the framework of the AQCM approach. These results
strongly point to an important role of $^{14}$C as a building block of cluster structures in medium
mass nuclei.

\begin{acknowledgments}
The numerical calculation has been performed using the computer facility of 
Yukawa Institute for Theoretical Physics,
Kyoto University. This work was supported by KAKENHI Grant Number 17K05440 of Japan,
and
U.S. Department of Energy, Office of Science, Office of
Nuclear Physics under Award No. DE-SC0013037.
\end{acknowledgments}

\bibliography{biblio_ni.bib}

\end{document}